\documentclass[12pt]{article}

\begin{document}

\newpage
\pagenumbering{arabic}
\thispagestyle{empty}

\begin{center}
 \bf \Large
Renormalization and dimensional regularization for a scalar field
with Gauss-Bonnet-type coupling to curvature
\end{center}

\begin{center}
{\bf    Yu.\,V. Pavlov{\,}\footnote{E-mail: \ pavlov@lpt.ipme.ru
}}  \\[4mm]
{\small \it
  Institute of Mechanical Engineering, Russian Academy of Sciences, \\
  61 Bolshoy, V.O., St.\,Petersburg, 199178, Russia  \\[1mm]
		     and     \\[1mm]
A.\,Friedmann Laboratory for Theoretical Physics,\\
St.\,Petersburg, Russia
}
\end{center}

\begin{abstract}
\noindent
{\bf Abstract.}
    We consider a scalar field with a Gauss-Bonnet-type coupling to
the curvature in a curved space-time.
    For such a quadratic coupling to the curvature, the metric
energy-momentum tensor does not contain derivatives of the metric
of orders greater than two.
    We obtain the metric energy-momentum tensor and find
the geometric structure of the first three counterterms to the
vacuum averages of the energy-momentum tensor for an arbitrary
background metric of an $N$-dimensional space-time.
    In a homogeneous isotropic space, we obtain the first three
counterterms of the $n$-wave procedure, which allow calculating
the renormalized values of the vacuum averages of the energy-momentum
tensors in the dimensions $N=4,5$.
    Using dimensional regularization, we establish  that the geometric
structures of the counterterms in the $n$-wave procedure coincide
with those in the effective action method.

\vspace{14pt}
\hfill 
\end{abstract}

\section{Introduction}
\hspace{\parindent}
    Quantum field effects in a curved space-time are intensively studied
nowadays and may be important for the cosmology of the early Universe
and for astrophysics (see~\cite{GMM}, \cite{BD}).
    Calculations in a curved space-time require describing
the interaction of the matter field with the external gravitational field.
    In the case of a scalar field, different types of coupling to
the curvature have been considered.
    The minimal coupling consists in replacing partial derivatives with
covariant ones in the free-field equations.
    If, moreover, we add the term $ \xi R \varphi $,
where $R$ is the scalar curvature, to the equation for the scalar
field $\varphi(x)$, then the equation for the massless field become
conformally invariat under a special choice  $\xi=\xi_c$.
    In several papers, the condition $\xi=\xi_c$ is regarded as
preferable (see, e.g.,~\cite{GribPMTr03}).
    For scalar fields in inflationary models~\cite{Linde},
the condition $\xi=0$ is usually assumed.
    In investigations of quantum effects in a curved space-time,
models with arbitrary values of $\xi$ are intensively
studied (see, e.g.,~\cite{BLMPvBMRHMPM00}).
    In the case of interacting fields, it is impossible to preserve
the conformal invariance not only effective action
but also an usual action after renormalization in a curved
space-time~\cite{BD}.
    The models with an arbitrary $\xi$, which are not conformally
invariant already at the classical level, can be generalized by adding
terms  quadratic in the curvature, and so on.
    But to avoid a radical change of the theory, we must preserve
the important property of the theories with a coupling of
the form $ \xi R \varphi $:
    the metric energy-momentum tensor (EMT) of the scalar field
with such a coupling does not contain derivatives of the metric of orders
greater than two.
    The presence of higher-order derivatives of the metric
in the metric EMT and therefore in the Einstein equations would result
in a qualitative change of the theory (see~\cite{Wald77})
already at the classical level!

    The requirement that higher-order derivatives of the metric be absent
was previously used as the basis for generalizing the theory of gravity
to higher dimensions~\cite{Lovelock71}.
    These conditions are satisfied in the multidimensional
Einstein-Gauss-Bonnet theories of gravity arising as low-energy
approximations of string theories~\cite{Zwiebach85}.
    In theories with the dilaton coupled to the curvature by
the Gauss-Bonnet invariant, the coupling of the scalar field to
the curvature is such that the EMT does not contain higher-order
derivatives of the metric.
    Such theories also arise as low-energy consequences in
string models (see, e.g.,~\cite{Ketov90}).

    In the present paper, we consider renormalization of the vacuum
averages of the EMT of a scalar field with a Gauss-Bonnet-type
coupling to the curvature,
find the geometric structures of the counterterms in an arbitrary metric,
and consider the $n$-wave procedure~\cite{ZlSt}, which is one of
the most effective ways to calculate the renormalized values of the EMT
in homogeneous isotropic spaces.
    In Sec.~2, we calculate the metric EMT and express it in terms
of the conformal Weyl tensor.
    This expression is convenient for calculations in
homogeneous isotropic spaces.
    In Sec.~3, we find the geometric structure of the counterterms to
the vacuum averages of the EMT in the effective action method.
    In Sec.~4, we quantize the scalar field in a homogeneous isotropic
space and calculate the vacuum averages of the EMT with respect to
the vacuum determined by the Hamiltonian diagonalization method.
    In Sec.~5, we calculate the first three counterterms of
the $n$-wave procedure in a homogeneous isotropic space-time and use
dimensional regularization to establish that their geometric structure
coincides with the structure of the counterterms in the effective action
method.
    In conclusion, we summarize the results in this work.
    In Appendix~A, we give expressions for the variations of some
geometric quantities as well as the Bianchi identities and their
consequences that are necessary for deriving the EMT and the counterterms
to its vacuum averages.
    In Appendix~B, we give the expressions for those geometric quantities
in an $N$-dimensional homogeneous isotropic space-time that are
used in this work.

    All calculations are done for the $N$-dimensional space-time,
that is necessary for dimensional regularization and
can be used to study higher-dimensional models.
    We use the system of units where $\hbar =c=1$.
    The signs of the curvature tensor and the Ricci tensor are
chosen such that
$ R^{\, i}_{\ jkl} = \partial_{\,l}  \Gamma^{\, i}_{\, jk} -
\partial_k  \Gamma^{\, i}_{\, jl} +
\Gamma^{\, i}_{\, nl} \Gamma^{\, n}_{\, jk} -
\Gamma^{\, i}_{\, nk} \Gamma^{\, n}_{\, jl}\ $
and  $\, R_{ik} = R^{\, l}_{\ ilk}$\,,
where $\Gamma^{\, i}_{\, jk}$ are Christoffel symbols.

\section{Scalar field with a Gauss-Bonnet coupling to the curvature}
\hspace{\parindent}
    We consider a complex scalar field  $\varphi(x)$ of mass $m$
with the Lagrangian
    \begin{equation}
L(x)=\sqrt{|g|} \left[\, g^{ik}\partial_i\varphi^*\partial_k \varphi -
(m^2 + V(R))\, \varphi^* \varphi \, \right]
\label{Lag}
\end{equation}
    and the corresponding equation of motion
\begin{equation}
 ({\nabla}^i {\nabla}_{\! i} + V(R) + m^2) \varphi(x)=0 \,,
\label{Eqm}
\end{equation}
   where ${\nabla}_{\! i}$ are the covariant derivatives in
the $N$-dimensional space-time with the metric $g_{ik}$,
$ g\!=\!{\rm det}(g_{ik})$,
and $\ V(R)$ denotes a function depending on
invariant combinations of the curvature tensor $R^{\,i}_{\ jkl}$
and the metrical tensor,
    \begin{equation}
V(R)=\xi R + \zeta R^2 + \kappa R_{ij}R^{ij} +
\chi R_{ijkl}R^{ijkl} + \ldots .
\label{VRR}
\end{equation}
     Equation~(\ref{Eqm}) is conformally invariant for $m=0$  and
$V(R) = \xi_c R $\,, where $ \xi_c = (N-2)/\,[4(N-1)] $ \
($\xi_c=1/6$ for $N=4$).
     The minimal coupling of the scalar field to the curvature
corresponds to $V(R)\equiv 0$.

    In general, nonzero constants with the dimension [mass]${}^{-2}$
at the terms quadratic in the curvature in~(\ref{VRR}) yield
third- and fourth-order derivatives of the metric in the metric EMT
and consequently in the Einstein equations.
    It is known (see~\cite{Wald77}) that such terms,
even with small coefficients, lead to a radical change of the theory.
    If we require that the metric EMT of the scalar field should
contain no derivatives of the metric of an order greater than two,
then the function
    \begin{equation}
V = \xi R + \zeta R_{GB}^{\,2} \,,  \ \ \  \mbox{where} \ \ \ \ \
R_{GB}^{\,2} \stackrel{\rm def}{=}
R_{lmpq} R^{\,lmpq} - 4 R_{lm} R^{\,lm} + R^2 \,,
\label{V}
\end{equation}
    can be taken as $V(R)$.
    For such a coupling of the scalar field to the curvature,
varying the action with respect to the metric and using formulas in
Appendix~A, we obtain the expression for the metric EMT
     \begin{eqnarray}
T_{ik} \! &\!=\!& \! \partial_i\varphi^* \partial_k\varphi+
\partial_k\varphi^*\partial_i\varphi - g_{ik} \partial^{\,l}\varphi^*
\partial_l\varphi + g_{ik} m^2 \varphi^* \varphi -
     \nonumber    \\
\! &\!-\!& \! 2 \xi \left( G_{ik} + \nabla_{\! i} \nabla_{\! k} -
g_{ik} \nabla^l \nabla_{\! l} \right) (\varphi^* \varphi)  -
2 \zeta \left( E_{ik} + P_{ik} \right) (\varphi^* \varphi) \,,
\label{TGB}
\end{eqnarray}
    where $G_{ik}= R_{ik} - R g_{ik}/2 $  is the Einstein tensor and
\begin{eqnarray}
E_{ik} = \frac{1}{ \sqrt{|g|} }\, \frac{\delta}{\delta g^{ik}}
\int \!  R_{GB}^{\,2} \, \sqrt{|g|} \, d^N x =
2 R_{ilmp} R_k^{\ lmp} -\frac{g_{ik}}{2} R_{lmpq} R^{\,lmpq} {}-
\nonumber        \\
{} - 4 R^{\, lm} R_{limk} -4 R_{il} R_k^{\,l} + 2 g_{ik} R_{lm} R^{\,lm}
+ 2 R R_{ik} - \frac{g_{ik}}{2} R^2 \,, \phantom{xx}
\label{Eik}
\end{eqnarray}
      \begin{eqnarray}
P_{ik} = 2 \, \biggl[ \, R \nabla_{\! i} \nabla_{\! k} +
2 R_{ik} \nabla_{\! l} \nabla^l + 2 g_{ik} R_{lm} \nabla^l \nabla^m -
R g_{ik} \nabla_{\! l} \nabla^l -
      \nonumber     \\
{} - 4 R_{l(i} \nabla_{\! k)} \nabla^l -
2 R_{ilkm} \nabla^l \nabla^m \biggr] \,.
\phantom{xxxx}     \label{Pik}
\end{eqnarray}
    Here, we introduce the notation
$A_{ n(i } B_{ k) }=( A_{ni} B_{k} + A_{nk} B_{i} )/2 $ \
    for the symmetrization.
    We note that \ $R_{GB}^{\, 2} \equiv 0$ \ for $ N=2,3 $; therefore,
no new effects can arise from $\zeta \ne 0$ in these dimensions.
    In the four-dimensional space-time,
we have \ $E_{ik}=0$\, (see~\cite{Lanczos38}),
but \ $P_{ik}(\varphi^*\varphi) \ne 0$ \ for a metric of the general
form and $\varphi(x) \ne {\rm const}$.

    The Gauss-Bonnet theorem holds in even-dimensional
spaces (see, for example,~\cite{TsFKS}):
for a compact oriented manifold $M$ with the even dimension $N=2k$,
the Euler characteristic, which is a topological invariant, is given by
       \begin{equation}
\chi(M) = \int_M E(x) \sqrt{ g } \, d^N x  \,,
\label{chi}
\end{equation}
    where $g_{ik}$ is Riemannian metric on $M$ and
       \begin{equation}
E(x) = \frac{(-1)^{k}}{(4\pi)^k k! 2^k}
\sum_{\pi, \sigma} (-1)^\pi (-1)^\sigma
R^{\pi(1) \pi(2)}_{\phantom{\pi(1) \pi(}\sigma(1) \sigma(2)} \ldots
R^{\pi(N-1) \pi(N)}_{\phantom{\pi(N-1) \pi(}\sigma(N-1) \sigma(N)} \,.
\label{Ex}
\end{equation}
    The summation is performed over all pairs $\{ \pi, \sigma \}$
of permutations of the set
$\{ 1, \ldots, N \}$,  and  $(-1)^\pi$  and  $ (-1)^\sigma $  are
the signs of the permutations.
    For $N=2$, we have \ $E(x) = - (4 \pi)^{-1} R(x) $.
    For $N=4$, the expression after the sum sign in~(\ref{Ex})
coincides with $R_{GB}^{\,2}$ (see~\cite{Lanczos38} for a proof).
    It is therefore natural to call the coupling of the scalar field
to the curvature in Eq.~(\ref{Eqm}) with $V(R)$ of form~(\ref{V})
a Gauss-Bonnet-type coupling.

    To investigate the EMT in a homogeneous isotropic space further,
we write the formulas for $E_{ik}$ and $P_{ik}$ expressing
the Riemann tensor in terms of the Weyl conformal tensor:
      \begin{equation}
C_{iklm} = R_{iklm} + \frac{2}{N\!-\!2} \biggl( R_{m\,[\,i}\,
g_{k\,]\,l} - R_{l\,[\,i}\, g_{k\,]\,m} \biggr) + \frac{2 \, R
\,g_{l\,[\,i}\,g_{k\,]\,m} }{(N\!-\!1)(N\!-\!2)} \,.
\label{Ciklm}
\end{equation}
    Here, the square brackets in the subscript denote antisymmetrization:
$A_{ n[i } B_{ k]m }=( A_{ni} B_{km} - A_{nk} B_{im} )/2 $. \
   Substituting (\ref{Ciklm}) in (\ref{Eik}) and (\ref{Pik}), we obtain
         \begin{equation}
E_{ik} = 2 C_{ilmp} C_k^{\ lmp} - \frac{g_{ik}}{2} C_{lmpq} C^{lmpq}
- (N - 4) {}^{(3)}\!H_{ik} \,,
\label{EikC}
\end{equation}
      \begin{eqnarray}
P_{ik} =  4 \left[ \, \frac{N\!-\!3}{N\!-\!2} \Biggl( \, R_{ik}
{\nabla}_{\! l} \nabla^l + g_{ik} R_{lm} {\nabla}^l \nabla^m -
2 R_{l(i} {\nabla}_{\! k)} \nabla^l \right. +
          \nonumber        \\
{} + \left. \frac{ N R }{2(N\!-\!1)} \left( {\nabla}_{\! i} \nabla_{\! k}
- g_{ik} {\nabla}_{\! l} {\nabla}^l \right) \Biggr)
-  C_{ilkm} {\nabla}^l {\nabla}^m \right],
\label{PikC}
\end{eqnarray}
    where the tensor
      \begin{eqnarray}
{}^{(3)}\!H_{ik} = \frac{4}{N-2} \,C_{ilkm} R^{\,lm} +
\frac{2(N-3)}{(N-2)^2} \left[ 2 R_{il} R_k^{\,l} -
\frac{N}{N-1}\, R R_{ik}  \right. -   \phantom{xx} \nonumber    \\
- \, \left. g_{ik} \biggl( R_{lm} R^{\,lm} -
\frac{N+2}{4(N-1)}\,R^2 \,\biggr) \right] \phantom{xx}
\label{3Hik}
\end{eqnarray}
    which was introduced in~\cite{Bunch79},
is covariantly conserved in the conformally flat case
(i.e., for ${C_{iklm}=0}$) and in a homogeneous isotropic space in
particular.

\section{Geometric structure of counterterms}

\hspace{\parindent}
    In a space-time with a metric of the general form, it is convenient
to use the dimensionally regularized effective action
to analyze the geometric structure of the divergences of
the vacuum averages of the EMT.
    For the complex scalar field $\varphi(x)$
with equation of motion~(\ref{Eqm}), the one-loop effective action
can be written as (see~\cite{BD}, \cite{Fulling})
       \begin{equation}
S_{eff} = \int L_{eff}(x) \sqrt{|g|}\, d^N x \,,
\label{Seff}
\end{equation}
    where
       \begin{equation}
L_{eff}(x) = (4 \pi)^{-N/2} \left( \frac{M}{m}
\right)^{2\varepsilon} \sum_{j=0}^\infty a_j(x) \, m^{N_0-2j}\,
\Gamma\biggl(j-\frac{N}{2}\biggr)  \,,
\label{Leff}
\end{equation}
       \begin{equation}
a_0(x)=1 \,, \ \ \ \ \
a_1(x) = \frac{1}{6}\,R - V = \left( \frac{4-N}{12 (N \!-\!1)} +
\Delta \xi \right) R - \zeta R_{GB}^{\,2} \,,
\label{a0a1}
\end{equation}
       \begin{eqnarray}
a_2(x) = \frac{1}{180} R_{lmpq} R^{\,lmpq} -
\frac{1}{180} R_{lm} R^{\,lm} + \frac{1}{72} R^2 -
\frac{1}{30} \nabla^l \nabla_{\!l} R -   \nonumber    \\
{} - \frac{1}{6} R V + \frac{1}{2} V^2 +
\frac{1}{6} \nabla^l \nabla_{\!l} V \,,
\label{a2}
\end{eqnarray}
    $N$ is the space-time dimension considered as a variable
analytically continued to the complex plane,
$\varepsilon$ is a complex parameter,
$M$ is a constant having the dimension of mass~\cite{tHooft} introduced
to restore the standard dimension (length)${}^{-N_0}$ of $L_{eff}$ in
the case $N=N_0-2\varepsilon$, $ \, \Gamma(z)$ is the gamma function,
and $  \Delta \xi \equiv \xi_c - \xi $.

     It follows from~(\ref{Ciklm}) that the expression for $a_2$
in the case  of Gauss-Bonnet-type coupling~(\ref{V}) to the curvature
can be written as
    \begin{eqnarray}
a_2(x) = \frac{(N\!-\!6)\, R_{GB}^{\,2}}{720\, (N\!-\!3)} +
\frac{(N\!-\!2) C_{lmpq} C^{\,lmpq}}{240\, (N - 3)} -
\frac{1}{6} \nabla^l \nabla_{\!l} \biggl( \frac{1}{5} R - V \biggr) +
\nonumber        \\
+ \left( \frac{ (N\!-\!4) (N\!-\!6)}{480\,(N\!-\!1)^2} - \frac{\Delta \xi
(N\!-\!4)}{12\,(N\!-\!1)} +\frac{(\Delta\xi)^2}{2} \right) R^2 +
\phantom{x}     \label{a2m}  \\
+ \left( \frac{(N\!-\!4)}{12(N\!-\!1)} - \Delta \xi  \right)
\zeta R R_{GB}^{\,2}  + \frac{1}{2}\, \zeta^2 R_{GB}^{\,4} \,.
\phantom{x}    \nonumber
\end{eqnarray}

    The first $[N_0/2]+1$ terms in~(\ref{Leff})
are excluded to obtain the renormalized Lagrangian $L_{eff}$
($[b]$ denotes the integer part of a number $b$).
   Varying the terms in the effective action corresponding to $j=0,1,2$
with respect to $g_{ik}$ and using formulas in Appendix~A,
we obtain the terms subtracted from the vacuum EMT\,:

     \begin{equation}
T_{ik,\varepsilon}[0]=- \frac{m^{N_0}}{2^{N_0} \pi^{N_0/2}} \left(
\frac{4 \pi M^2}{m^2} \right)^{\! \displaystyle \varepsilon }
\Gamma  \biggl( \varepsilon - \frac{N_0}{2} \biggr) \, g_{ik} \,,
\label{TE0}
\end{equation}

    \begin{eqnarray}
T_{ik,\varepsilon}[1] &=& \frac{m^{N_0-2}}{2^{N_0-1} \pi^{N_0/2}}
\left( \frac{4 \pi M^2}{m^2} \right)^{\! \displaystyle \varepsilon }
\Gamma \biggl( 1 -  \frac{N}{2} \biggr) \left[
\left( \frac{1}{6} - \xi \right) G_{ik} -\zeta E_{ik} \right] =
\nonumber     \\
&=& \frac{m^{N_0-2}}{2^{N_0-1} \pi^{N_0/2}}
\left(\frac{4 \pi M^2}{m^2} \! \right)^{\! \displaystyle \varepsilon }
\Biggl[\, \Delta \xi\, \Gamma \biggl(1 - \frac{N}{2} \biggr) G_{ik} -
\label{TE1}       \\
&&{}  -  \frac{\Gamma \left(3 - \frac{N}{2} \right)}{(N-2)}
\Biggl(  \frac{G_{ik}}{3 (N-1)} + \zeta \frac{ 4E_{ik}}{(N-4)} \Biggr)
\, \Biggr] \,,   \nonumber
\end{eqnarray}

      \begin{eqnarray}
T_{ik,\varepsilon}[2] \!=\!
\frac{m^{N_0-4}}{(4\pi)^{\frac{N_0}{2}}}
\!\left( \!\frac{4 \pi M^2}{m^2}\!\right)^{\! \displaystyle \varepsilon }
\!\left\{ \frac{\Gamma \left( 2 - \frac{N}{2}\right)}{360 (N \!-\! 3) }
\biggl[ (N\!-\!6) E_{ik} + 3 (N\!-\!2) W_{ik} \biggr] \right. +
 \nonumber                \\
{} + \Biggl[ \, \frac{\Gamma \left( 4 - \frac{N}{2}\right)}{60
\, (N - 1)^2 } + \Delta \xi\, \frac{\Gamma \left(3 - \frac{N}{2}
\right)}{3\, (N - 1) } + (\Delta \xi)^2  \, \Gamma \biggl( 2 -
\frac{N}{2} \biggr) \,\Biggr] {}^{(1)}\! H_{ik} -
\nonumber           \\
{} - \zeta \, \frac{\Gamma \left(3\!-\!\frac{N}{2}\right)}{3(N\!-\!1)}
\Biggl[ \biggl( R_{ik} + {\nabla}_{\! i} {\nabla}_{\! k} -
g_{ik} {\nabla}_{\! l}{\nabla}^l \biggr) R_{GB}^{\, 2}  +
\biggl( E_{ik} + P_{ik} \biggr) R
\Biggr] +
      \label{TE2ik}     \\
{}+ \zeta \, 2 \, \Gamma \left( 2 - \frac{N}{2}\right)
\Biggl[ \, \zeta\, \biggl( \frac{g_{ik}}{4} R_{GB}^{\, 2} + E_{ik} +
P_{ik} \biggr) \, R_{GB}^{\, 2} -
   \nonumber           \\
{} - \Delta \xi \, \Biggl( \biggl(R_{ik} + {\nabla}_{\! i} {\nabla}_{\! k}
- g_{ik} {\nabla}_{\! l}{\nabla}^l \biggr) R_{GB}^{\, 2}  +
\biggl( E_{ik} + P_{ik} \biggr) R \Biggr) \Biggr] \Biggr\} \,,
\nonumber
\end{eqnarray}
    where

    \begin{equation}
{}^{(1)}\! H_{ik}= \frac{\delta  {\displaystyle \int \!  R^2  \,
\sqrt{|g|} \, d^N x}} {\sqrt{|g|} \,  \delta g^{ik}} = 2 \biggl(
\nabla_{\! i} \nabla_{\! k} R - g_{ik} \nabla^l \nabla_{\! l} R
\biggr) + 2 R \biggl( R_{ik} -\frac{1}{4} R \, g_{ik} \biggr),
\label{1Hik}
\end{equation}

\vspace{10mm}
    \begin{eqnarray}
W_{ik} = \frac{1}{\sqrt{|g|}}\, \frac{\delta}{ \delta g^{ik}}
\int \!  C_{lmpq} C^{\,lmpq} \sqrt{|g|} \, d^N x =
  \phantom{xxxxxxx}
\label{Wik}    \\
{} = \, E_{ik} + \, \frac{4(N-3)}{N-2} \Biggl( 2 R^{\, lm} R_{limk} -
\frac{ g_{ik} }{ 2 } R_{lm} R^{\, lm} -
\frac{ N\, R R_{ik} }{ 2(N - 1) } {}+ \phantom{xx}
\nonumber        \\
{} + \frac{ N\, R^2 g_{ik} }{ 8(N - 1) }
+ \frac{N-2}{2(N-1)} \nabla_{\! i} \nabla_{\! k} R +
\frac{ g_{ik} }{ 2(N-1) } \nabla^l \nabla_{\! l} R -
\nabla^l \nabla_{\! l} R_{ik} \Biggr).
\nonumber
\end{eqnarray}
     In the conformally flat case, we have
$ C_{iklm}\!=0 $,\ \ $E_{ik}/(4\!-\!N)\!={}^{(3)}\!H_{ik} $, and
$W_{ik}\!=0$.
    Hence,
    \begin{eqnarray}
T_{ik,\varepsilon}[1] =
\frac{m^{N_0-2}}{2^{N_0-1} \pi^{N_0/2}}
\left(\frac{4 \pi M^2}{m^2} \! \right)^{\! \displaystyle \varepsilon }
\Biggl[\, \Delta \xi\, \Gamma \biggl(1 - \frac{N}{2} \biggr) G_{ik} +
\nonumber     \\
{} + \frac{\Gamma \left(3 - \frac{N}{2} \right)}{(N-2)}
\Biggl( \, \zeta 4 \, {}^{(3)}\!H_{ik} - \frac{G_{ik}}{3 (N-1)}
\Biggr) \, \Biggr] \,,
\label{TE1kp}
\end{eqnarray}
      \begin{eqnarray}
T_{ik,\varepsilon}[2] = \frac{m^{N_0-4}}{(4\pi)^{N_0/2}}
\left( \frac{4 \pi M^2}{m^2}\right)^{\! \displaystyle \varepsilon }
\left\{ {}^{(3)}\! H_{ik} \left[
\frac{ - \Gamma \left(4 -\frac{N}{2}\right) }{ 90\, (N-3) } \right. +
\right.
\phantom{xxxxx}    \label{TE2kp}      \\
{} + \left. \zeta 4 \Gamma \left(3 -\frac{N}{2}\right)\!
\left( \Biggl( \frac{N-4}{12(N\!-\!1)} - \Delta \xi \Biggr) R +
\zeta R_{GB}^{\,2} \right) \right] +
\phantom{xxxx}    \nonumber           \\
{} +  {}^{(1)}\! H_{ik} \Biggl[ \,
\frac{\Gamma \left( 4 - \frac{N}{2}\right)}{60\, (N - 1)^2 } +
\Delta \xi\, \frac{\Gamma \left(3 - \frac{N}{2} \right)}{3\, (N - 1) }
+ (\Delta \xi)^2  \, \Gamma \biggl( 2 - \frac{N}{2} \biggr) \,\Biggr] -
\phantom{xx}     \nonumber      \\
{} - \zeta \, \frac{\Gamma \left(3\!-\!\frac{N}{2}\right)}{3(N\!-\!1)}
\Biggl[ \biggl( R_{ik} + {\nabla}_{\! i} {\nabla}_{\! k} -
g_{ik} {\nabla}_{\! l}{\nabla}^l \biggr) R_{GB}^{\, 2}  + P_{ik} R
\Biggr] + \zeta \, 2 \, \Gamma \left( 2 \!-\! \frac{N}{2}\right)
\times      \hspace{-4mm}    \nonumber    \\
{} \times \Biggl[ \zeta \biggl( \frac{g_{ik}}{4} R_{GB}^{\, 2} +
P_{ik} \biggr) R_{GB}^{\, 2} -
\Delta \xi \Biggl( \biggl(R_{ik} + {\nabla}_{\! i} {\nabla}_{\! k}
- g_{ik} {\nabla}_{\! l}{\nabla}^l \biggr) R_{GB}^{\, 2} +
P_{ik} R \Biggr) \Biggr] \Biggr\}. \hspace{-5mm}
\nonumber
\end{eqnarray}

     Assuming that the vacuum averages $\langle \, T_{ik}\, \rangle$
of the EMT are sources of the gravitational field
(see~\cite{GMM}, \cite{BD}), i.e.,
     \begin{equation}
G_{ik} + \Lambda g_{ik} = - 8 \pi G
\biggl( T_{ik}^{\,b} + \langle \, T_{ik}\, \rangle \biggr)\,,
\label{Eeq}
\end{equation}
   where $\Lambda$ and $G$ are the cosmological and gravitational
constants and  $ T_{ik}^{\,b} $  is the EMT of the background matter,
we can draw the following conclusion from
formulas (\ref{TE0})--(\ref{TE2ik}).
     The first three subtractions from the vacuum EMT in
the $N$-dimensional space-time correspond to renormalizing
the cosmological and gravitational constants and
the parameters at the terms of the second, third, and fourth orders
in the curvature in the bare gravitational Lagrangian of the form
     \begin{equation}
L_{gr,\,\varepsilon} = \sqrt{|g|} \biggl[
\frac{ R \!-\! 2 \Lambda_\varepsilon }{16 \pi G_\varepsilon} +
\alpha_\varepsilon  R_{GB}^{\,2} +
\beta_\varepsilon   R^2 +
\gamma_\varepsilon  C_{lmpq} C^{lmpq} +
\delta_\varepsilon  R R_{GB}^{\,2} +
\theta_\varepsilon  R_{GB}^{\,4}  \biggr].
\label{Lgr0}
\end{equation}
      By (\ref{TE0}), subtracting $ T_{ik,\varepsilon}[0] $
corresponds to an infinite renormalization of
the cosmological constant $\Lambda_\varepsilon$.
    It follows from~(\ref{TE1}) that subtracting $ T_{ik,\varepsilon}[1] $
corresponds to a renormalization of the gravitational
constant $G_\varepsilon$ (finite for $N \to 4$ and $\xi=\xi_c$)
and an infinite renormalization of the parameter $\alpha_\varepsilon$
(for $\zeta \ne 0$).
    By~(\ref{a2m}) and (\ref{TE2ik}), subtracting $ T_{ik,\varepsilon}[2] $
corresponds to a renormalization of the parameters
$\alpha_\varepsilon, \beta_\varepsilon, \gamma_\varepsilon,
\delta_\varepsilon$, and $ \theta_\varepsilon$.
    For $\xi=\xi_c$ and $N \to 4$, the renormalization of the parameters
$\beta_\varepsilon$ and $\delta_\varepsilon$ is finite.

    We note that the form of the terms $\alpha_\varepsilon  R_{GB}^{\,2}$
in~(\ref{Lgr0})  agrees with use of dimensional regularization
($E_{ik} \equiv 0$ only for integer $N=2,3,4$).

    For $N\to 4$, the products $E_{ik}\Gamma (1 - (N/2))$ and
$E_{ik}\Gamma (2 - (N/2))$
have finite limits for an arbitrary space-time metric
because the dependence of the expressions for $E_{ik}$ on $N$ is assumed
to be rational in the analytic continuation with respect to the dimension,
$\, E_{ik}=0$ \, for $N=4$, and the gamma function
has first-order poles at the points $0$ and $-1$.
    The corresponding terms in~(\ref{TE1}) and (\ref{TE2ik})
are therefore finite, and it is unnecessary to subtract them to obtain
finite quantities in Eq.~(\ref{Eeq}), describing the back reaction of
the quantized field on the metric.
    But without such subtractions, the effective action remains
divergent, and the expression for the anomalous trace of the vacuum EMT
differs from the standard one even for $\zeta=0$.
    Because the corresponding terms arise for different methods of
regularization (see~\cite{GMM}, \cite{BD}),
it is customary to retain them in the counterterms to the vacuum EMT.

    The questions whether finite renormalizations are necessary and
what the values of the renormalized parameters are pertain to experiment.
    As noted in~\cite{GMM}, the renormalized parameters at non-Einstein
terms in the gravitational Lagrangian are possibly equal to zero.

\section{The scalar field in a homogeneous isotropic space}
\hspace{\parindent}
    We write the metric of an $N$-dimensional homogeneous isotropic
space-time in the form
    \begin{equation}
ds^2=g_{ik}dx^i dx^k = a^2(\eta)\,(d{\eta}^2 - d l^2) \,,
\label{gik}
\end{equation}
    where $ d l^2=\gamma_{\alpha \beta}\, d x^\alpha d x^\beta $
is the metric of the $(N-1)$-dimensional space of constant curvature
$K=0, \pm 1 $.
    The complete set of solutions of Eq.~(\ref{Eqm}) in metric~(\ref{gik})
can be found in the form
    \begin{equation}
\varphi(x) = a^{-(N-2)/2} (\eta)\, g_\lambda (\eta) \Phi_J ({\bf x}) \,,
\label{fgf}
\end{equation}
	where
    \begin{equation}
g_\lambda''(\eta)+\Omega^2(\eta)\,g_\lambda(\eta)=0 \,,
\label{gdd}
\end{equation}
       \begin{equation}
\Omega^2(\eta)=m^2 a^2 +\lambda^2 - \Delta \xi\,a^2 R +
\zeta a^2 R_{GB}^{\,2} \,,
\label{Ome}
\end{equation}
     \begin{equation}
\Delta_{N-1}\,\Phi_J ({\bf x}) = - \Biggl( \lambda^2 -
\biggl( \frac{N-2}{2} \biggr)^2 K \Biggr) \Phi_J  ({\bf x})\,,
\label{DFlF}
\end{equation}
    the prime denotes the derivative with respect to the conformal
time~$\eta$,
  and $J$ is the set of indices (quantum numbers) labeling
the eigenfunctions of the Laplace-Beltrami operator $\Delta_{N-1}$
in the ($N-1$)-dimensional space.
    We note that because the eigenvalues of the operator $-\Delta_{N-1} $
are nonnegative, we have the inequality
$ \lambda^2-((N-2)/\,2)^2\,K \ge 0 $\,.

    According to the Hamiltonian diagonalization method~\cite{GMM}
(see~\cite{PvIJA} for the case of an arbitrary $V(R)$),
the functions $g_\lambda(\eta)$ should satisfy the initial conditions
    \begin{equation}
g_\lambda'(\eta_0)=i\, \Omega(\eta_0)\, g_\lambda(\eta_0) \,, \ \
\ |g_\lambda(\eta_0)|= \Omega^{-1/2}(\eta_0)\,.
\label{icg}
\end{equation}

     For quantizing, we decompose the field $ \varphi(x) $ with respect to
the complete set of solutions of~(\ref{fgf}),
    \begin{equation}
\varphi(x)=\int \! d\mu(J)\,\biggl[ \varphi^{(+)}_J \,a^{(+)}_J +
\varphi^{(-)}_J \, a^{(-)}_J \,\biggr] \ ,
\label{fff}
\end{equation}
    where
    \begin{equation}
\varphi^{(+)}_J (x)=\frac{a^{-(N-2)/2} (\eta)}{\sqrt{2}}\,
g_\lambda(\eta)\,\Phi^*_J({\bf x}) \ , \ \  \ \ \varphi^{(-)}_J(x) =
\biggl(\varphi_J^{(+)}(x) \biggr)^* \ ,
\label{fpm}
\end{equation}
     and impose the commutation relations
\begin{equation}
\left[a_J^{(-)}, \ \stackrel{*}{a}\!{\!}_{J'}^{(+)}\right] =
\left[\stackrel{*}{a}\!{\!}_J^{(-)}, \ a_{J'}^{(+)}\right] =
\delta_{JJ'} \ , \ \ \
\left[a_J^{(\pm)}, \ a_{J'}^{(\pm)}\right] =
\left[\stackrel{*}{a}\!{\!}_J^{(\pm)}, \
\stackrel{*}{a}\!{\!}_{J'}^{(\pm)} \right]=0 \,.
\label{aar}
\end{equation}
     It is convenient to express the averages of the EMT operator for
the vacuum $| 0\rangle $  annihilated by the operators
$a_J^{(-)}$ and $\ \stackrel{*}{a}\!{\!}_{J}^{(-)}$
in terms of the bilinear combinations
    \begin{equation}
S=\frac{|g_\lambda'|^2 + \Omega^2\,|g_\lambda|^2}{4 \, \Omega}
-\frac{1}{2} \,, \ \ U=\frac{\Omega^2 \, |g_\lambda|^2-
|g_\lambda'|^2}{2\, \Omega} \,  \,, \ \ V= - \frac{d (g_\lambda^*
g_\lambda)}{2\, d \eta}
\label{SUV}
\end{equation}
    of the functions~$g_\lambda $ and~$g_\lambda^*$,
which satisfy the system of differential equations
    \begin{equation}
S'= \frac{\Omega'}{2\, \Omega} \, U \ , \ \ \ U'=
\frac{\Omega'}{\Omega} \, (1+ 2 S) - 2 \, \Omega V \ , \ \ \ V'= 2
\,\Omega \, U
\label{sdu}
\end{equation}
    in accordance with~(\ref{gdd}).
    Taking the initial conditions
$ S(\eta_0)=U(\eta_0)=V(\eta_0)=0 $ \, following from~(\ref{icg})
into account, we can rewrite Eqs.~(\ref{sdu}) as a system of the Volterra
integral equations
     \begin{equation}
U(\eta) + i V(\eta) = \int_{\eta_0}^{\, \eta} \! w(\eta_1)\, (1+2
S(\eta_1)) \exp[ 2\,i\,\Theta(\eta_1,\eta)]\,d\eta_1  \,,
\label{ie1}
\end{equation}
    \begin{equation}
S(\eta)=\frac{1}{2}\,\int_{\eta_0}^{\, \eta} \! d\eta_1 \,
w(\eta_1)\, \int_{\eta_0}^{\, \eta_1} \! d\eta_2 \,w(\eta_2)\,
(1+2 S(\eta_2)) \cos[2\,\Theta(\eta_2,\eta_1)]  \,,
\label{ie2}
\end{equation}
      where
$$
w(\eta) =\frac{\Omega'(\eta)}{\Omega(\eta)} \ , \ \ \ \ \
\Theta(\eta_1, \eta_2) = \int_{\eta_1}^{\eta_2} \Omega(\eta)\,d\eta \,.
$$

    To obtain the vacuum EMT, we use the following summation formulas
for the eigenfunctions of the operator
$\Delta_{N-1}$ (see~\cite{Pv3}):
     \begin{equation}
\hspace{-25pt} \sum_{\phantom{xxxxx} J\,
{\scriptscriptstyle (\lambda=const)}} \hspace{-19pt} |\Phi_J({\bf x})|^2
= f(\lambda) \,,
\label{fF}
\end{equation}
     \begin{equation}
\hspace*{-25pt} \sum_{\phantom{xxxxx} J\,
{\scriptscriptstyle (\lambda=const)}} \hspace{-22pt}
\left[ \left( \partial_\alpha \Phi^*_J \right) \partial_\beta \Phi_J +
\left( \partial_\beta \Phi^*_J \right) \partial_\alpha \Phi_J \right] =
\frac{ 2 \gamma_{\alpha \beta }}{N\!-\!1} \left( \lambda^2 \!-\!
\biggl(\frac{N\!-\!2}{2} \biggr)^2 K \right) f(\lambda)  \,,
\label{dfF}
\end{equation}
    \begin{equation}
\hspace*{-25pt} \sum_{\phantom{xxxx} J\, {\scriptscriptstyle
(\lambda=const)}} \hspace{-22pt}
\left[\left(\tilde{\nabla}_{\!\alpha} \tilde{\nabla}_{\!\beta}
\Phi^*_J \right) \Phi_J + \Phi^*_J \tilde{\nabla}_{\!\alpha}
\tilde{\nabla}_{\!\beta} \Phi_J \right] = - \, \frac{2
\gamma_{\alpha \beta }}{N\!-\!1} \left(\!\lambda^2\! -\!
\biggl(\frac{N\!-\!2}{2} \biggr)^2 K\! \right) f(\lambda) \,,
\label{nfF}
\end{equation}
    where the sign  $\sum$ denotes integration for continuous $J$ and
$\tilde{\nabla}_{\!\alpha}$ is the covariant derivations in the
$(N-1)$-dimensional space with the metric $\gamma_{\alpha \beta}$. \
    In quasi-Euclidean case $(K=0)$, the function $f(\lambda)$ is given
by
     \begin{equation}
f(\lambda) = \frac{B_N}{2} \lambda^{N-2}\ , \ \ \ \mbox{} \ \ \ \ \
B_N = \left[ 2^{N-3} \pi^{\frac{N-1}{2}}\,
\Gamma\,\biggl( \frac{N\!-\!1}{2} \biggr) \right]^{-1}
\label{fBN}
\end{equation}
(see~\cite{Pv3} for the case $ K \neq 0 $).

    Substituting decomposition (\ref{fff}) in (\ref{TGB}) and using
(\ref{EikC})--(\ref{3Hik}), (\ref{aar}), (\ref{SUV}),
(\ref{fF})--(\ref{nfF}), and formulas in Appendix~B,
we obtain the (divergent) expressions
    \begin{equation}
\langle 0 |\,T_{ik}| 0\rangle =
\frac{B_N}{a^{N-2}} \int \! d\mu(\lambda) \, \tau_{ik}
\label{Ttik}
\end{equation}
    for the vacuum averages of the EMT, where
           \begin{eqnarray}
\tau_{00}=\Omega\, \biggl( S+\frac{1}{2}\biggr) + (N -1)
\biggl( \Delta \xi - \tilde{\zeta}\, (c^2 + K) \biggr) \times
\nonumber       \\
{} \times \left[\, c V + \biggl( c'+(N - 2) c^2 \biggr)\,
\frac{1}{\Omega} \left( S+\frac{1}{2} U + \frac{1}{2} \right) \right],
\label{ts00}
\end{eqnarray}
    \begin{eqnarray}
\tau_{\alpha \beta} = \gamma_{\alpha \beta} \, \Biggl\{\,
\frac{1}{(N \!-\! 1) \Omega}  \left[ \lambda^2 \left( S +
\frac{1}{2} \right) - \left( \Omega^2 - \lambda^2 \right)
\frac{U}{2} \, \right] -
\phantom{xxxxxxx}         \nonumber   \\
{}- 2 \biggl( \Delta \xi \!- \tilde{\zeta} (c^2 \!+\! K) \biggr) \Omega U
+ \biggl[ \Delta \xi (N\!-\!1) \!- \tilde{\zeta} \biggl( (N\!+\!1)
(c^2\!+\!K) \!- 2 c'\biggr) \biggr] c V \!-
\nonumber       \\
{} - \frac{1}{\Omega} \left( S+\frac{1}{2} U + \frac{1}{2} \right)
 \Biggl[ \Delta \xi \biggl( (N\!-\!1) c' + (N\!-\!2)K \biggr) -
\phantom{xxxxxxx}  \label{tsab}         \\
{} - \tilde{\zeta} \Biggl( (N\!-\!1)c^2 \biggl( 3c'\!-\! 2
(c^2\!+\!K) \biggr) + K \biggl( (N\!+\!1)c' +
(N\!-\!4) (c^2\!+\!K) \biggr) \Biggr) \Biggr] \Biggr\}\,,
\nonumber
\end{eqnarray}
  $ \tilde{\zeta} \equiv \zeta a^{-2}\, 2 (N\!-\!2)(N\!-\!3)$, \,
and \, $c\equiv \! a'/a$.

     In the four-dimensional space-time, the integration measure
in expression~(\ref{Ttik}) has the form (see~\cite{GMM})
    \begin{equation}
\int\!d\mu(\lambda)\ldots = \left\{
\begin{array}{ll}
{\ \displaystyle \int_0^\infty d \lambda\, \lambda^{2} } \ldots, \
\
&  K=0, -1 \,,   \\[11pt]
{\displaystyle \sum \limits_{\lambda=1}^\infty \lambda^2 } \ldots,
&  K=1.
\end{array}             \right.
\label{mera}
\end{equation}
     In the $N$-dimensional case, we have \,
$d\mu(\lambda) = \sigma(\lambda) d \lambda $  \,
(see~\cite{Pv3}), where
     \begin{equation}
\sigma(\lambda)= \lambda^{N-2} + \alpha_N \lambda^{N-4} + \beta_N
\lambda^{N-6} + \ldots \,,
\label{sig}
\end{equation}
      \begin{equation}
\alpha_N = - \frac{1}{24}\,(N-2)(N-3)(N-4) K \,,
\label{aNW}
\end{equation}
      \begin{equation}
\beta_N = \frac{1}{5760} \, (N-2)(N-3)(N-4)(N-5)(N-6)(5N-8) K^2 \,.
\label{bNW}
\end{equation}

\section{The $n$-wave procedure}

\hspace{\parindent}
    The $n$-wave procedure proposed in~\cite{ZlSt} is often used
to calculate the renormalized vacuum EMT in homogeneous isotropic spaces.
    For the $N$-dimensional homogeneous isotropic space-time,
the $n$-wave procedure can be expressed by the formulas
(see~\cite{Pv3}):
     \begin{equation}
\langle 0|\,T_{ik}| 0\rangle_{ren}=\frac{B_N}{a^{N-2}}
\lim_{\Lambda \to \infty} \left[ \int \limits^{\Lambda} \!
d\mu(\lambda) \, \tau_{ik} - \sum \limits_{l=0}^{[N/2]} \int
\limits_0^{\Lambda} \! d \lambda \lambda^{N-2} \, a_{ik}[l] \right],
\label{Trik}
\end{equation}
     where
\begin{equation}
a_{ik}[l] = \frac{1}{l!} \lim_{n \to \infty} \frac{\partial^{\,
l}}{\partial (n^{-2})^l} \left( \frac{ \tau_{ik} (n \lambda, n m)
\sigma(n \lambda)}{n^{N-1} \lambda^{N-2}}  \right).
\label{aikl}
\end{equation}

     To find\, $a_{ik}[l]$\, explicitly, we expand\,
$S,U$, and $V$ (see~(\ref{SUV})) in inverse powers of $n$ after replacing
$\lambda \to n \lambda$\, and $ \, m \to n m $\, for
$ n \to \infty $:  $ S=\sum_{k=1}^\infty n^{-k} S_k, \ \ldots $.
     Using consecutive iterations in integral
equations~(\ref{ie1}) and (\ref{ie2}) and the stationary phase method,
we obtain the first nonzero expansion terms
     \begin{equation}
V_1=W \,,\ \ U_2=DW \,,\ \ S_2=\frac{1}{4} W^2 \,, \ \
V_3=\frac{1}{2} W^3 - D^2 W - \frac{\omega}{2}
D\left(\frac{q}{\omega^3} \right),
\label{V1U2S2}
\end{equation}
           \begin{equation}
U_4=\frac{3}{2} \, W^2 DW - D^3 W - D \left( \frac{\omega}{2} \, D
\biggl( \frac{q}{\omega^3} \biggr) \right) + \frac{q}{2\omega^2} DW \,,
\label{U4}
\end{equation}
           \begin{equation}
S_4=\frac{3}{16} \, W^4 + \frac{1}{4}\,(D W)^2 - \frac{1}{2}\, W
D^2 W -\frac{1}{4}\,\omega W D\biggl(\frac{q}{\omega^3}\biggr) \,,
\label{S4}
\end{equation}
    where
\begin{equation}
q = \!\left( \Delta \xi R - \zeta R_{GB}^{\, 2} \right)\! a^2 , \ \
\omega=(m^2a^2+\lambda^2)^{1/2} , \ \ W=\frac{\omega\,'}{2 \omega^2} \,,
\ \ D=\frac{1}{2 \omega} \, \frac{d}{d\eta} \,.
\label{qoWD}
\end{equation}
    We note that the terms that are nonlocal in time
(i.e., depend on both\, $\eta$ and $\eta_0$) are excluded from
expressions (\ref{V1U2S2})--(\ref{S4}).
    Such terms are absent if\,
$V_1(\eta_0) = V_3(\eta_0) = U_2(\eta_0) = U_4(\eta_0) = 0 $,
which is assumed below.
    In particular, there are no nonlocal terms if
the first $2[N/2]$ derivatives of the scale factor $a(\eta)$
of the metric vanish at the initial instant.

    Using~(\ref{ts00}), (\ref{tsab}), (\ref{sig}), and
(\ref{V1U2S2})--(\ref{S4}), we obtain the expressions for $a_{ik}[l]$,

     \begin{equation}
a_{00}[0] = \tau_{00}[0] = \frac{\omega}{2} \,, \ \ \ \ a_{\alpha
\beta}[0] = \tau_{\alpha \beta}[0] = \gamma_{\alpha \beta}\,
\frac{\lambda^2}{2 (N-1) \, \omega} \ ,
\label{at00}
\end{equation}

    \begin{equation}
a_{ik}[1]=\tau_{ik}[1] + \frac{\alpha_N}{\lambda^2} \tau_{ik}[0]
\,, \ \ \ \ a_{ik}[2]=\tau_{ik}[2] + \frac{\alpha_N}{\lambda^2}
\tau_{ik}[1] + \frac{\beta_N}{\lambda^4} \tau_{ik}[0] \,,
\label{at12}
\end{equation}
	 where

    \begin{eqnarray}
\tau_{00}[1]=\omega \, S_2 + (N-1) \biggl( \Delta \xi -
\tilde{\zeta} (c^2+K) \biggr) c V_1 +
\phantom{xxxx}   \nonumber      \\
{} + \frac{N\!-\!1}{8 \omega} \biggl[ \,\Delta \xi \, 2 (N\!-\!2) \,
(c^2\!-\!K) + \tilde{\zeta} (c^2\!+\!K)
\left( (4\!-\!3N) c^2 +(N\!-\!4) K \right) \biggr] ,
\label{t001}
\end{eqnarray}

                      \begin{eqnarray}
\tau_{\alpha \beta}[1] = \gamma_{\alpha \beta} \left\{
\frac{1}{(N\!-\!1)\omega} \! \left[ \lambda^2 S_2 - \frac{m^2 a^2}{2}
\biggl( U_2 +\frac{q}{2 \omega^2} \biggr) \right]  -
2 \Delta \xi \, \omega U_2 \right. +
\phantom{xx}        \nonumber     \\
{} + 2 \tilde{\zeta} (c^2\!+\!K) \, \omega U_2 +
\biggl[ \Delta \xi (N\!-\!1) - \tilde{\zeta} \biggl(
(N\!+\!1) (c^2\!+\!K) - 2 c'\biggr) \biggr] c V_1 +
\nonumber       \\
{} + \frac{1}{4\omega} \Biggl[\, \Delta \xi (N\!-\!2)
\left( c^2 - K - 2c'\right) +
\tilde{\zeta} \, \Biggl( \, c^2 (3N\!-\!4)
\biggl( 2c' - \frac{3}{2} c^2 \biggr) +
\label{tab1}        \\
{}+ \left. K \biggl( 2 N c' - 3 N c^2 + \frac{3}{2} (N\!-\!4) K
\biggr) \Biggr) \Biggr] \right\},
\phantom{xx}    \nonumber
\end{eqnarray}

             \begin{eqnarray}
\tau_{00}[2] = \omega\, \biggl( S_4+\frac{q}{4 \omega^2}\,U_2 +
\frac{q^2}{16 \omega^4} \biggr) +
(N\!-\!1) \biggl( \Delta \xi - \tilde{\zeta} (c^2\!+\!K) \biggr) c V_3 +
\nonumber    \\
{} +  \biggl[ \, \Delta \xi \, 2 (N\!-\!2)
(c^2 \!-\! K) + \tilde{\zeta} (c^2\!+\!K) \biggl( (N \!-\! 4) K -
(3 N \!-\! 4) c^2 \biggr) \biggr] \times
\label{t002}           \\
{} \times \frac{N \!-\! 1}{4\, \omega}
\biggl( S_2+\frac{1}{2}\,U_2+\frac{q}{4\omega^2} \biggr) \,,
\phantom{xxxxxxx}     \nonumber
\end{eqnarray}

                      \begin{eqnarray}
\tau_{\alpha \beta}[2] = \gamma_{\alpha \beta} \left\{ \!
\frac{1}{N \!-\! 1} \Biggl[ \frac{\lambda^2}{\omega} \Biggl( S_4 +
\frac{q U_2}{4 \omega^2} + \frac{q^2}{16 \omega^4} \Biggr) \! -
\frac{m^2 a^2}{2 \,\omega}  \Biggl( U_4 +\frac{q^2}{4 \omega^4} +
\frac{q S_2}{\omega^2} \Biggr) \Biggr] + \right. \hspace{-4mm}
      \nonumber   \\
{}+ \biggl[ \Delta \xi (N \!-\!1) - \tilde{\zeta} \biggl(
(N\!+\!1) (c^2\!+\!K) - 2 c'\biggr) \biggr] c V_3
- 2 \biggl( \Delta \xi -  \tilde{\zeta} (c^2\!+\!K) \biggr) \times
\hspace{-4mm}        \nonumber       \\
\times \Biggl( \omega U_4 \!- \frac{q U_2}{2 \omega} \Biggr) +
\Biggl[ \Delta \xi (N\!-\!2) ( c^2 \!-\! K \!-\! 2c')
+ \tilde{\zeta} \, \Biggl( c^2 (3N\!-\!4)
\biggl( 2c' - \frac{3}{2} c^2 \biggr) \! +
\hspace{-4mm}        \nonumber     \\
{}+ \left. K \biggl( 2 N c' - 3 N c^2 + \frac{3}{2} (N\!-\!4) K
\biggr) \Biggr) \Biggr] \, \frac{1}{2\omega} \,
\Biggl( S_2+\frac{1}{2}\,U_2+\frac{q}{4\omega^2} \Biggr) \right\}.
\phantom{xxxx}    \label{tab2}
\end{eqnarray}
    The above expressions exhaust all subtractions in the dimensions $N=4,5$.

    The vacuum EMT renormalized according to~(\ref{Trik})  is
covariantly conserved.
    This is proved using expressions~(\ref{at00}) and (\ref{at12})
and the equalities
$\nabla^i (\tau_{ik}/a^{N-2})=0 $ and $\nabla^i (\tau_{ik}[l]/a^{N-2})=0 $
following from~(\ref{sdu}), (\ref{ts00}), (\ref{tsab}), and
(\ref{t001})--(\ref{tab2}).

    As in~\cite{MMSH}, we perform the dimensional regularization
to clarify the geometric structure of the counterterms of
the $n$-wave procedure.
    To calculate the integrals in the dimensionally regularized
counterterms
    \begin{equation}
T_{ik, \varepsilon }[l]=\frac{B_N}{a^{N-2}} (M)^{2 \varepsilon}
\int_{0}^{\infty} \! d\lambda \,\lambda^{N-2} a_{ik, \varepsilon}[l] \,,
\label{kTik}
\end{equation}
    where $a_{ik,\varepsilon}[l] $ are defined by formulas
(\ref{at00})--(\ref{tab2}) with replacement $N \to N_0 - 2\varepsilon $,
we use the equality
    \begin{equation}
\int_0^\infty  x^k\, (1+x^2)^{-p}\, dx =\frac{\Gamma\,(
\frac{k+1}{2})\,\Gamma\,(p-\frac{k+1}{2})}{2\, \Gamma\,(p)} \,.
\label{iGf}
\end{equation}
    If the integral in the left-hand side of~(\ref{iGf}) does not
exist in the usual sense, then it is assumed to be equal to
the analytic continuation of the right-hand side of~(\ref{iGf}) to
the corresponding values of\, $k$ and\, $p$.
    As a result of calculations that are cumbersome
but contain no principal difficulties, we obtain the respective
expressions~(\ref{TE0}), (\ref{TE1kp}), and (\ref{TE2kp}) for
zeroth, first, and second counterterms in the $n$-wave procedure
by taking formulas in Appendix~B into account.
    The geometric structure of the first three subtractions
in the $n$-wave procedure therefore coincides with that in
the effective action method.

\section{Conclusion}
\hspace{\parindent}
    In this paper, we consider a scalar field with a Gauss-Bonnet-type
coupling to the curvature such that the metric EMT contains no
derivatives of the metric of order greater than two.
    We obtain expressions~(\ref{TGB})--(\ref{Pik}) for the EMT and
represented them in terms of the Weyl conformal tensor.
    This representation is convenient for calculations in both
conformally flat (in particular, homogeneous isotropic) spaces
and Ricci-flat (i.e., for $R_{ik}=0$) spaces
(see~(\ref{EikC}) and (\ref{PikC})).
    We found the geometric structure of the first three counterterms to
the vacuum EMT in an arbitrary metric for an $N$-dimensional space-time
(see~(\ref{TE0})--(\ref{TE2ik})).
    In the dimensions $N=4,5$, these counterterms exhaust all subtractions.
    The analysis of the geometric structure of these counterterms allows
concluding that the first three subtractions correspond
to renormalizing the cosmological and gravitational constants
and the parameters at the terms of the second, third, and fourth order in
the curvature in the bare gravitational Lagrangian of form~(\ref{Lgr0}).
    For a homogeneous isotropic space-time, we obtained
formulas~(\ref{Ttik})--(\ref{tsab}), which determine the nonrenormalized
vacuum averages of the EMT.
    We found the first three counterterms of  $n$-wave
procedure~(\ref{Trik}) for the vacuum EMT (see~(\ref{at00})--(\ref{tab2})).
    Using dimensional regularization, we established that
the geometric structure of the counterterms of the $n$-wave procedure
in a homogeneous isotropic space coincides with the structure of
counterterms~(\ref{TE0}), (\ref{TE1kp}), (\ref{TE2kp}) obtained in
the effective action method.
    Formulas~(\ref{Ttik})--(\ref{tsab}), (\ref{Trik}), and
(\ref{at00})--(\ref{tab2}) allow to calculating the renormalized vacuum
averages of the EMT for a scalar field with a Gauss-Bonnet-type
coupling to the curvature in a homogeneous isotropic space-time
in the dimensions $N=4,5$.

    Taking a possible coupling of the scalar field to
the Gauss-Bonnet invariant $R_{GB}^{\,2}$ into account may be important
for the early Universe.
    A nonzero value of parameter $\zeta$ in the equations of the scalar
field  may influence black-hole evaporation
(see~\cite{MignemiStewart93} for the case of a dilaton coupled to
the Gauss-Bonnet invariant), the parameters of the so-called bosonic
stars (see, e.g.,~\cite{MielkeS}), and so on.
    The questions concerning the value of the parameter $\zeta$ and
the parameter $\xi$ ultimately pertain to experiment.

\newpage
\noindent
{\bf \Large        Appendix\ A  }

\setcounter{equation}{0}
\renewcommand{\theequation}{A.\arabic{equation}}

\vspace{2mm}
    Here, we give expressions for the variations of some geometric
quantities as well as the Bianchi identities and their consequences that
are necessary for obtaining the EMT of a scalar field with
a Gauss-Bonnet-type coupling to the curvature and the counterterms to
its vacuum averages.
    The semicolon in the indices denotes the corresponding
covariant derivatives.

    For calculating variations, we introduce the notation
$ \delta g_{ik} = h_{ik} $ \ and \ $ h=h^i_i $ \
for the variations of the metric.
    We then have $ \delta g^{ik} = - h^{ik} $\,,
    \begin{equation}
\delta \sqrt{|g|} = - \frac{1}{2} \sqrt{|g|}\, g_{ik} \delta g^{ik}
=  \frac{1}{2} \sqrt{|g|}\, h  \,, \ \ \
\delta \Gamma^{\, i}_{\, kl} = \frac{1}{2} \left(
h^i_{k;\, l} + h^i_{l;\, k} - h_{kl}{}^{;\, i}  \right),
\label{dgikdgdGikl}
\end{equation}
    \begin{equation}
\delta R^{\, i}_{\ jkl} =
\frac{1}{2} \left( h^i_{j;\, kl} + h^i_{k;\, jl} - h_{jk}{}^{;\, i}{}_l
- h^i_{l;\, jk} - h^i_{j;\, lk} + h_{jl}{}^{;\, i}{}_k   \right),
\label{dRvikl}
\end{equation}
    \begin{equation}
\delta R_{\, ik}= \frac{1}{2} \left( h_{;\, ik} +
h_{ik}{}^{;\, l}{}_l - h^l_{i;\, kl} - h^l_{k;\, il} \right),
\label{dRik}
\end{equation}
    \begin{equation}
\delta R = h_{;\, l}{}^l - h_{lm}{}^{;\, lm} - R^{\, lm} h_{lm} \ ,
\label{dR}
\end{equation}
    \begin{equation}
\delta (R^2) = 2 R \, h_{;\, l}{}^l - 2 R \, h_{lm}{}^{;\, lm} -
2 R R^{\, lm} h_{lm} \ ,
\label{dR2}
\end{equation}
    \begin{equation}
\delta (R_{\,ik} R^{\, ik}) = R^{\, ik} \left( h_{;\, ik} +
h_{ik}{}^{;\, l}{}_l - h^l_{i;\, kl} - h^l_{k;\, il} \right)
- 2 R_{\,il}R^{\, l}_k h^{ik} \,,
\label{dRik2}
\end{equation}
    \begin{equation}
\delta (R_{lmpq} R^{\, lmpq}) = 2 R^{\, lmpq} \left( h_{lp;\, mq} +
h_{lp;\, qm} \right) - 2 R_{lmpq} R_{\,i}{}^{mpq} h^{li} \,,
\label{dRlmpq2}
\end{equation}
    \begin{eqnarray}
\delta  ( R_{lmpq} R^{\,lmpq} - 4 R_{lm} R^{\,lm} + R^2 ) =
\phantom{xxxxxxxxxxx} \nonumber  \\
{} = 2 \left[\, ( 4 R_{\,il} R_{\, k}^{\,l} - R_{\,ilpq} R_k{}^{lpq} -
R R_{\,ik}) h^{ik} +  R_{\,i}{}^l{}_k{}^q ( h^{ik}{}_{;\, lq}
+ h^{ik}{}_{;\, ql} )  \right. +
\label{dGB}
\\
+ \left. R \, h_{;\, l}{}^l - R \, h^{ik}{}_{;\, ik} -
2 R^{\,lm} h_{;\, lm} - 2 R_{\, ik} h^{ik}{}_{;\, l}{}^l +
4  R_{\, il} h^{ik; \, l}{}_k \, \right].    \phantom{xx} \nonumber
\end{eqnarray}

    The Bianchi identities for a symmetric connection compatible with
the metric have the form
    \begin{equation}
R^{\, i}_{\ jkl;\, m} + R^{\, i}_{\ jmk;\, l} + R^{\, i}_{\ jlm;\, k} = 0
\,.
\label{Bianchi}
\end{equation}
    It follows from (\ref{Bianchi}) and the symmetry properties of
the curvature tensor that
    \begin{equation}
R^{\, n}{}_{\! ilm;\, n} = R_{\,im;\, l} - R_{\, il; \, m} \ , \ \ \ \
R^{\, l}_{\, m;\, l} = \frac{1}{2} R_{\,;\, m}  \ .
\label{Bianchi12}
\end{equation}
    Repeated differentiation of (\ref{Bianchi12}) yields the identities
    \begin{equation}
R^{\, ilkn}{}_{;\, ln} =
R^{\, ik}{}_{;\, l}{}^{l} - R^{\, il}{}^{;\, k}{}_{l} \,, \ \ \ \
R^{\, lm}{}_{;\, lm} = \frac{1}{2} R_{\,;\, m}{}^m  \ .
\label{Bianchi34}
\end{equation}
      Using (\ref{Bianchi12}) and identity
    \begin{equation}
( \nabla_{\!l} \nabla_{\!k} - \nabla_{\!k} \nabla_{\!l} ) A^{il} =
R^{\,i}{}_{nkl} A^{nl} - R_{\,nk} A^{in} \,,
\label{DlDAilsym}
\end{equation}
    which holds for a symmetric connection and an arbitrary tensor
of rank two, we obtain the consequence of the Bianchi identities
    \begin{equation}
R^{\,il}{}_{;\, kl} = \frac{1}{2} R^{\,;\, i}{}_k +
R^{\,i}{}_{mkl} R^{ml} - R^{\,im} R_{mk} \ .
\label{BianchiSl5}
\end{equation}

\vspace{4mm}
\noindent
{\Large \bf Appendix\ B }

\setcounter{equation}{0}
\renewcommand{\theequation}{B.\arabic{equation}}

\vspace{2mm}
    Here, we give expressions for some geometric quantities in the
$N$-dimensional homogeneous isotropic space-time with metric~(\ref{gik}).

     The Christoffel symbols are
     \begin{equation}
\Gamma^{\,0}_{\, 00}= \frac{a'}{a} \equiv c   \ ,
\ \ \ \  \Gamma^{\,i}_{\, 0 j}=c \, \delta^i_j   \ ,
\ \ \ \ \Gamma^{\,0}_{\, \alpha \beta}= c \, \gamma_{\alpha \beta} \ ,
\ \ \ \ \Gamma^{\,\alpha}_{\, \beta \delta}(g_{ik}) =
\Gamma^{\,\alpha}_{\, \beta \delta}(\gamma_{\nu \mu}) \,.
\label{GGG}
\end{equation}
    The nonzero components of the Ricci tensor and the scalar curvature are
     \begin{equation}
R_{00} = (N-1) \, c'   \ , \ \ \ \ R_{\alpha
\beta}=-\gamma_{\alpha \beta} \left[ \, c'+(N-2) (c^2+K) \right],
\label{R00Rab}
\end{equation}
     \begin{equation}
R= a^{-2}(N-1) \left[ \, 2c'+(N-2)(c^2+K) \right].
\label{RRRR}
\end{equation}
    The components of the Einstein tensor are
     \begin{equation}
G_{00} =\! -\frac{(N\!-\!1) (N\!-\!2)}{2} (c^2\!+\!K) ,  \ \
G_{\alpha \beta}=\! \gamma_{\alpha \beta} (N\!-\!2) \left[
c'+\frac{(N\!-\!3)}{2} (c^2\!+\!K)\right].
\label{G00Gab}
\end{equation}
    Using (\ref{R00Rab}), (\ref{RRRR}) and the formula
       \begin{equation}
R_{lmpq} R^{\,lmpq} - 4 R_{lm} R^{\,lm} + R^2 =   C_{lmpq} C^{lmpq}
- \frac{4(N\!-\!3)}{N\!-\!2} \biggl( R_{lm} R^{\,lm} -
\frac{N\,R^2}{4(N\!-\!1)} \biggr),
\label{RRRRRC}
\end{equation}
    we find that
       \begin{eqnarray}
R_{lmpq} R^{\,lmpq} - 4 R_{lm} R^{\,lm} + R^2 &=&
a^{-4} (N\!-\!1) (N\!-\!2) (N\!-\!3) (c^2 +K) \times
\nonumber        \\
&\times& \left[\, 4 c'+ (N\!-\!4) \, (c^2 +K) \,\right]
\label{RGBoi}
\end{eqnarray}
    in the homogeneous isotropic case.
    The components of ${}^{(1)}\! H_{ik} $ and ${}^{(3)}\! H_{ik} $ are
given by
       \begin{eqnarray}
{}^{(1)}\! H_{00}= \frac{(N\!-\!1)^2}{a^2} \left[ 2 c'{\,}^2 - 4
c''c - 4 (N\!-\!4) c' c^2 - \frac{c^4}{2} (N\!-\!2) (N\!-\!10)
\right.  -
     \nonumber     \\
-\left. K(N-2) \left( \frac{N-2}{2}K + (N-6) c^2 \right) \right],
\phantom{xxxxx} \label{c1H00}
\end{eqnarray}
       \begin{eqnarray}
{}^{(1)}\! H_{\alpha \beta} = \gamma_{\alpha \beta} \, a^{-2}
(N-1) \, \biggl\{ 4 c^{(3)} + 4\, ( 2 N - 9 )\, c''c + 2\, (3 N -
11) \, c'{\,}^2 +
\nonumber           \\
+\, 6\, ( N^2 - 10 N + 20 ) \, c' c^2 + \frac{1}{2} ( N-2 ) ( N^2
- 15 N + 50 ) \, c^4 +
\nonumber    \\
+\, K(N\!-\!2)\, \biggl[ (N\!-\!5)(N\!-\!6) c^2 + 2(N\!-\!6) c' +
\frac{1}{2} (N\!-\!2)(N\!-\!5) K \, \biggr] \biggr\} ,
\phantom{xx} \label{c1Hab}
\end{eqnarray}
     \begin{equation}
{}^{(3)}\! H_{00} = a^{-2}\,2^{-1} (N-1) (N-2) (N-3) \, (c^2 + K)^2  \ ,
\label{c3H00}
\end{equation}
     \begin{equation}
{}^{(3)}\! H_{\alpha \beta} = - \gamma_{\alpha \beta}\, a^{-2} \,
2^{-1} (N-2) (N-3) \left[ 4 c' (c^2 +K) + (N-5) \, (c^2 +K)^2 \right].
\label{c3Hab}
\end{equation}

\vspace{11mm}
 {\bf \large Acknowledgments.}
    The author is grateful to Professor A.\,A.\,Grib and the participants
in the seminar of A.\,Friedmann Laboratory for Theoretical Physics
for helpful discussion.

    This work was supported by the Ministry of Education of
the Russian Federation  (Grant No. E02-3.1-198).

\newpage

\end{document}